\documentclass[11pt]{article}

\usepackage{float,graphicx}
\usepackage{subfigure}

\begin{document}
%
\title{Degree-correlation, robustness, and vulnerability in finite scale-free networks}


\author{ Jeremy F.~Alm\\ Department of Mathematics\\ Illinois College\\ Jacksonville, IL\\ \and
 Keenan M.~L.~Mack\\Department of Biology\\  Illinois College\\Jacksonville, IL}

\maketitle

\begin{abstract}
Many naturally occurring networks have a power-law degree distribution as well as a non-zero degree correlation.  Despite this, most studies analyzing the robustness to random node-deletion and vulnerability to targeted node-deletion have concentrated only on power-law degree distribution and ignored degree correlation.  This study looks specifically at the effect degree-correlation has on robustness and vulnerability in scale-free networks.  Our results confirm Newman's finding that positive degree-correlation increases robustness and  decreases vulnerability. However, we found that  networks with positive degree-correlation are more vulnerable to random node-deletion than to targeted deletion methods that utilize knowledge of initial node-degree only.  Targeted deletion sufficiently alters the topology of the network to render this method less effective than uniform random methods unless changes in topology are accounted for.  This result indicates the importance of degree correlation in certain network applications.
\end{abstract}


\section{Introduction}

Since the seminal paper of Barabasi and Albert \cite{Barabasi99} showing how networks built through growth and the preferential attachment of nodes naturally have a power-law degree distribution, work in network science has focused on networks following a power-law distribution instead of a binomial distribution.  Most real-world networks have been found to follow a power-law distribution \cite{Newman}.  In this study, we test the robustness to random node-deletion and vulnerability to targeted node-deletion in randomly-generated networks that not only follow a power-law degree distribution, but also have varying levels of degree-correlation.  Even though many naturally occurring networks have non-zero degree correlation, few of the previous studies looking at robustness and vulnerability of networks have included this important aspect of realism \cite{Britton10,Chen08,Gong13,Hasegawa11,Hebert13,Shams14,Zeng13} (however, see \cite{Hu12}). For example, in \cite{Bollobas, Flaxman}, the authors study the effect of node-deletion, both random and targeted, on the sizes of the connected components of power-law networks; they do not consider the effect of degree correlation.  

Random node-deletion models random failure in, say, a power grid.  Targeted node-deletion models sabotage in a power grid or computer network.  Naturally, one expects that targeted deletion would be more effective than random deletion at breaking a network into small connected components, and this is usually true.  However, in the presence of sufficiently high positive degree-correlation, if the targeted deletion is based on limited information --- only the degrees of nodes in the initial network --- targeted deletion performs worse than random deletion at breaking the network into small connected components.  We demonstrate this by deleting nodes from various 1000-node networks until all connected components have size at most 20, thus destroying the giant component.



\section{Methods}

One hundred different 1000-node networks were constructed via preferential attachment.  For each such network, a rewiring algorithm was applied to copies of the network to produce degree-correlation coefficients of -0.3, -0.2, -0.1, 0.0, 0.1, 0.2, and 0.3 (see Section \ref{subsec:rewiring}).  Three further copies of each rewired network were then created and subjected to three different node-deletion methods, whereby we deleted nodes until the largest connected component had at most 20 nodes:

\begin{itemize}
    \item[Method 1] (Random) Choose nodes to delete uniformly at random, until no connected component of the network has size exceeding 20
    \item[Method 2] (Initial Degree Rank) Delete nodes in decreasing order of node-degrees in the original network, until no connected component of the network has size exceeding 20
    \item[Method 3] (Updated Degree Rank -- Omniscient) Delete nodes in decreasing order of node-degree, recalculating the degree of every node every 10 deletions, until no connected component of the network has size exceeding 20
\end{itemize}

One may think of these three methods as ``no knowledge'', ``knowledge of initial degree rank'', and ``complete knowledge of the network structure'', respectively.  The results of the analysis are displayed in Figure \ref{fig:result}.

In order to estimate the variability  in our samples, we performed bootstrap resampling by taking 1,000,000 samples of size 100 (with replacement) from our original data sets for Methods 1 and 2, $r=0.3$.    


\subsection{Network generation} \label{subsec:nx}

We used the Python library \texttt{networkx}  to generate our networks.  Specifically, we used the function \texttt{barabasi\_albert\_graph(n, m)} with 1000 nodes ($n=1000$) and with each added node attaching to three existing nodes ($m=3$).


\subsection{A rewiring algorithm} \label{subsec:rewiring}

We consider the algorithm proposed in \cite{xulvi2004reshuffling} to give an existing network positive or negative degree correlation without changing the degree distribution. 

\begin{enumerate}
    \item Choose two edges $e_1,e_2$ uniformly at random, rejecting pairs of edges incident to a common node. Suppose the nodes are $n_1,n_2,n_3,n_4$ with $\deg(n_1)\leq \deg(n_2)\leq \deg(n_3)\leq \deg(n_4)$.
    \item \begin{enumerate}
        \item (positive correlation) Check to see that $n_1$ is not adjacent to $n_2$, and that $n_3$ is not adjacent to $n_4$; if so, delete  $e_1, e_2$ and add $n_1n_2$ and  $n_3n_4$ to the network; else proceed.
        
        \item (negative  correlation) Check to see that $n_1$ is not adjacent to $n_4$, and that $n_2$ is not adjacent to $n_3$; if so, delete  $e_1, e_2$ and add $n_1n_4$ and  $n_2n_3$ to the network; else proceed.

    \end{enumerate}
    \item Repeat a prescribed number of times.
\end{enumerate}


\section{Results}

\begin{figure}[H]
    \centering
     \subfigure[]{%
            \includegraphics[width=4.2in]{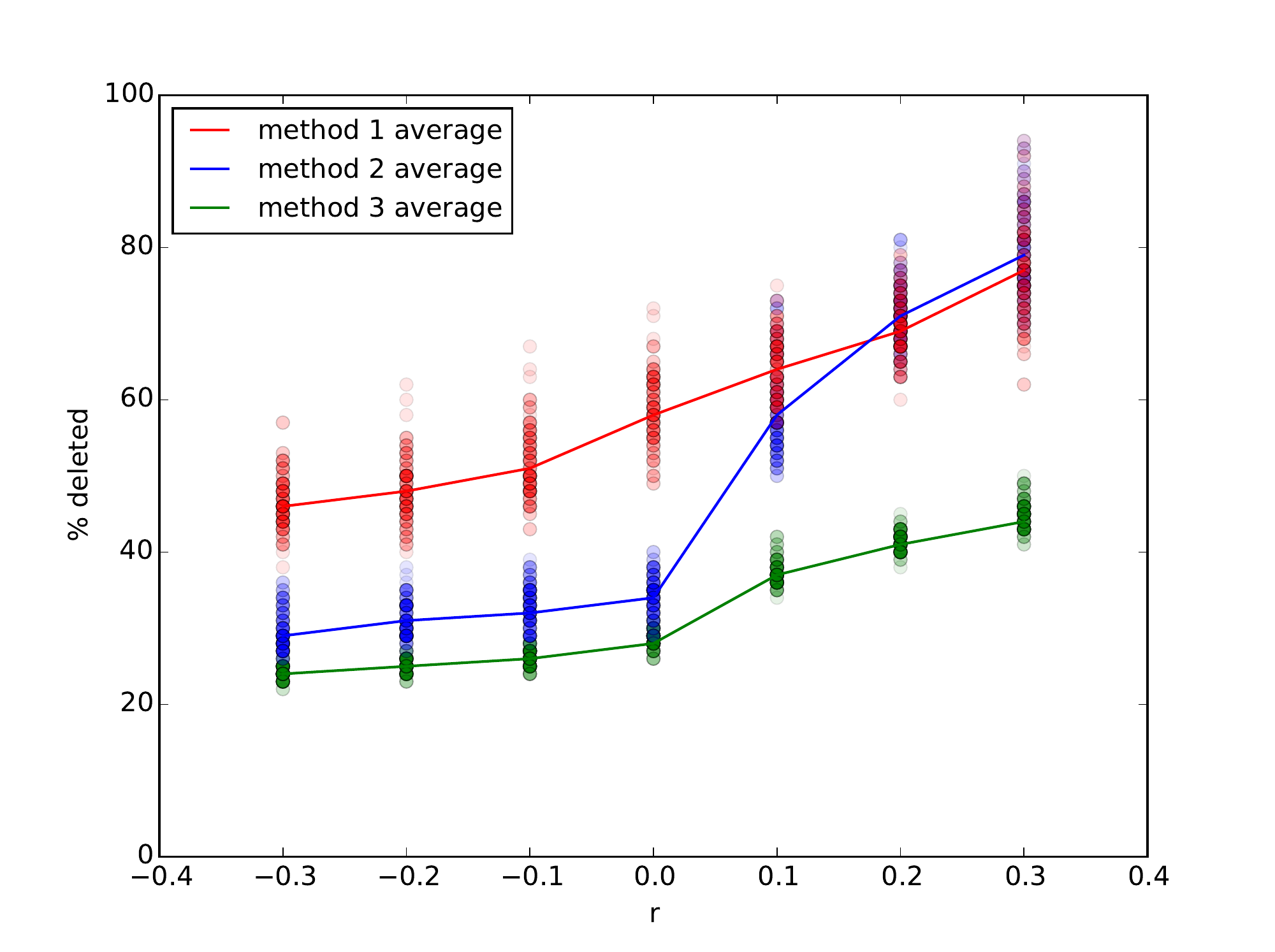}
             
            }
     \subfigure[]{%
     \includegraphics[width=4.2in]{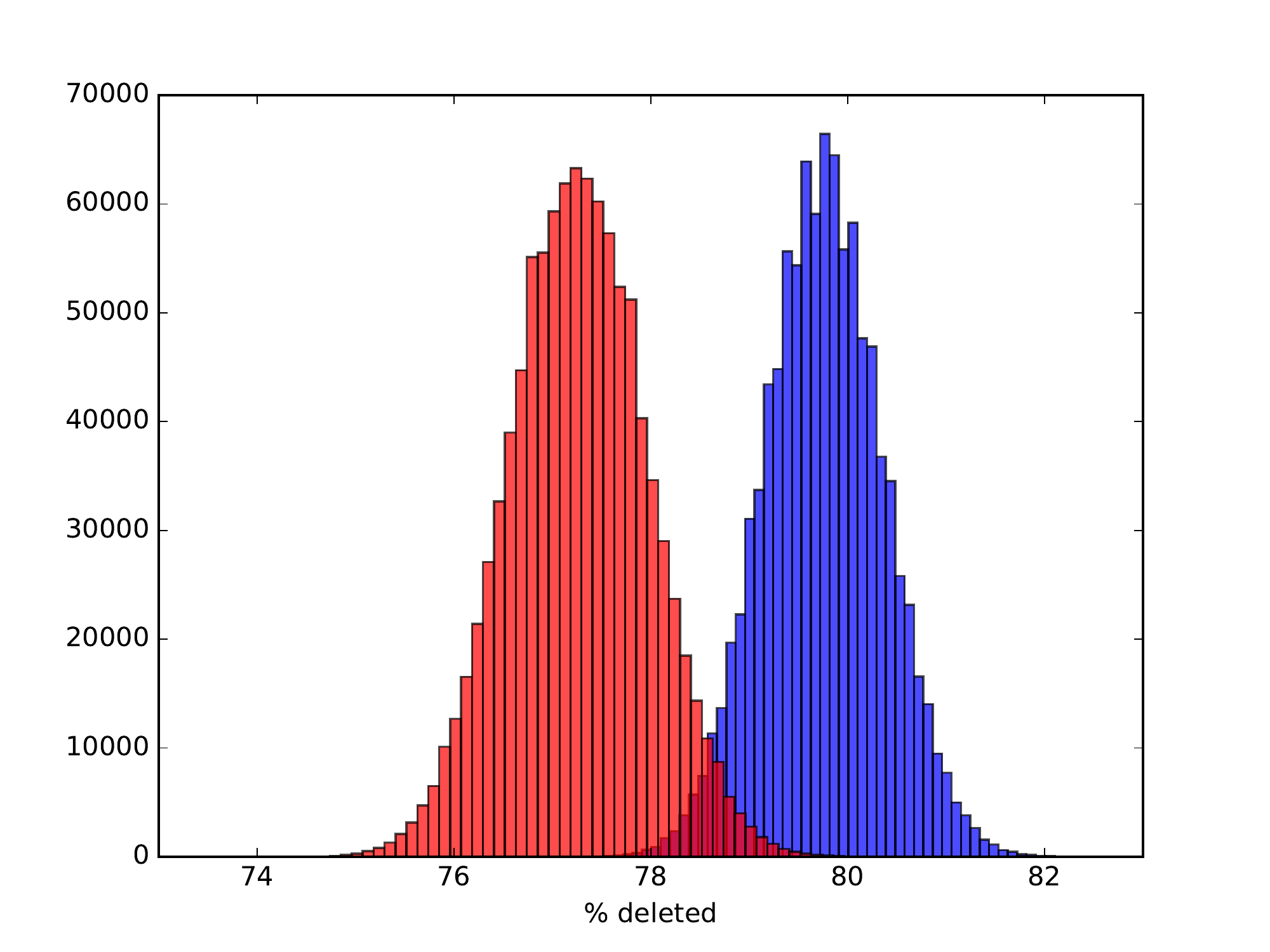}
     
            }%
            
            \caption{(a) Data and averages for different correlation levels. 
            (b) Resampling histogram for Method 1 (in red) and Method 2 (in blue) for $r=0.3$ with $10^6$ resamples }
           
    \label{fig:result}
\end{figure}

The percent deleted necessary to reduce the size of the largest connected component to 20 or fewer for each level of manipulated degree-correlation for each of the node-deletion methods is plotted in Figure \ref{fig:result} (a).  Histograms for the averages of each bootstrap resampling from our original data sets for Methods 1 and 2, $r=0.3$ are in Figure \ref{fig:result} (b).   The corresponding resampling intervals are non-overlapping at the 94.5\% confidence level.


\section{Discussion}

As expected, positive degree correlation raises the number of nodes which must be deleted in order to reduce the size of the largest connected component to 20 or fewer nodes \cite{Newman02,Hu12}.  Not surprisingly, under certain conditions, using knowledge of the initial degree rank (method 2) to inform the node-deletion method results in fewer nodes needing to be deleted than in a random node-deletion method (method 1), as previously found \cite{Bollobas, Flaxman}.  However, under sufficiently high positive degree-correlation, deleting nodes based on initial degree rank (method 2) actually requires more node-deletion than the random method (method 1).  In networks with sufficiently high positive degree correlation, in order for knowledge of initial degree rank to be effectively utilized, the degree rank must be continually updated as nodes are removed from the network (method 3).  Hu and Tang \cite{Hu12} showed previously that updating the degree rank does result in fewer nodes needing to be deleted than method 2, and the difference is greater in networks with positive degree correlation than those with negative degree correlation; however, due to the narrow scope of their chosen correlation coefficients (0.0579 and -0.0441), they did not discover the extent to which this performance difference is manifest.  Additionally, they did not compare the performance of the initial degree rank method to a random method, thus missing the important observation that method 2 can actually perform worse than method 1 in networks with sufficiently high positive correlation. It is worth noting that studies analyzing the effectiveness of various node-deletion methods most often use the default preferential attachment model, which has a degree-correlation of zero;  according to our results, this happens to be the condition under which there is the largest difference in performance between method 1 and method 2.  See Figure \ref{fig:result} (a).

Besides applications to power grids and computer networks, a possible implication of this result is for immunization protocols. Imagine a disease with, for simplicity, perfect transmission efficiency such that any individual within a population that shares a transmission vector with an infected individual will always become infected.  Therefore, in the network model, individuals are represented as nodes with the edges representing direct transmission vectors; thus all nodes with a connected path to an infected node also become infected.  If the network is connected, that means that if one node is infected, then the entire network will be infected.  Immunization can then be modeled by deletion of nodes from the network.  When enough individuals are immunized---that is, when enough nodes are deleted---the network becomes disconnected and part of the network will become isolated from the infection, effectively stopping the spread of the disease.  As immunization progresses through time, the network is decomposed into more and more, smaller and smaller connected components.  Each of these connected components, if harboring an infected individual, will become entirely infected, but the disease will not spread  to the other components.  Thus, the size of the largest connected component represents the worst-case scenario for the outbreak of the disease. (Setting the cutoff for an acceptable number of people to be infected is arbitrary and does not qualitatively affect our results.)  Therefore, this suggests that in human interaction networks, which often have a power-law degree distribution and positive degree correlation, having some knowledge of the network structure beyond initial degree rank could be critical for choosing the most effective immunization protocol. 

\section{Conclusion}
We found that  networks with positive degree-correlation are more vulnerable to random node-deletion than to targeted deletion methods that utilize knowledge of initial node-degree only.  Targeted deletion sufficiently alters the topology of the network to render this method less effective than uniform random methods unless changes in topology are accounted for.  This result indicates the importance of degree correlation in certain network applications.







\end{document}